%% file: camera_ready.tex
\documentclass[sigconf]{acmart}

\AtBeginDocument{%
  }

\copyrightyear{2025}
\acmYear{2025}
\setcopyright{acmlicensed}
\acmConference[MMAsia '25]{ACM Multimedia Asia}{December 9--12, 2025}{Kuala Lumpur, Malaysia}
\acmBooktitle{ACM Multimedia Asia (MMAsia '25), December 9--12, 2025, Kuala Lumpur, Malaysia}
\acmDOI{10.1145/3743093.3771012}
\acmISBN{979-8-4007-2005-5/2025/12}

\acmSubmissionID{260}
\usepackage{enumitem}

\begin{document}

\title{VSpeechLM: A Visual Speech Language Model for \\Visual Text-to-Speech Task}

\author{Yuyue Wang}
\authornote{Equal contributions.}
\email{wangyuyue123@ruc.edu.cn}
\orcid{0009-0005-6987-1028}
\affiliation{
  \institution{Gaoling School of Artificial Intelligence, Renmin University of China \country{}}
}

\author{Xin Cheng}
\email{chengxin000@ruc.edu.cn}
\orcid{0009-0001-7581-8662}
\affiliation{
  \institution{Gaoling School of Artificial Intelligence, Renmin University of China \country{}}
}
\authornotemark[1]

\author{Yihan Wu}
\email{yihanwu@ruc.edu.cn}
\orcid{0009-0001-0312-782X}
\affiliation{
  \institution{Gaoling School of Artificial Intelligence, Renmin University of China \country{}}
}

\author{Xihua Wang}
\email{xihuaw@ruc.edu.cn}
\orcid{0009-0002-9454-5775}
\affiliation{
  \institution{Gaoling School of Artificial Intelligence, Renmin University of China \country{}}
}

\author{Jinchuan Tian}
\email{jinchuat@andrew.cmu.edu}
\orcid{0000-0002-2129-471X}
\affiliation{
  \institution{Carnegie Mellon University \country{}}
}

\author{Ruihua Song}
\authornote{Corresponding author.}
\email{songruihua_bloon@outlook.com}
\orcid{0000-0001-6036-9035}
\affiliation{
  \institution{Gaoling School of Artificial Intelligence, Renmin University of China \country{}}
}

\input{00-abstract}


\maketitle

\input{01-intro}

\input{02-related}
\input{03-method}
\input{04-exp}
\input{05-conclusion}

\clearpage

\begin{acks}
This work is supported by the National Natural Science Foundation of China (No. 62276268).
\end{acks}

\bibliographystyle{ACM-Reference-Format}
\bibliography{ref}

\end{document}

%% file: 00-abstract.tex
\begin{abstract}
  The task of Visual Text-to-Speech (VisualTTS), also known as video dubbing, aims to generate speech synchronized with the lip movements in an input video, in additional to being consistent with the content of input text and cloning the timbre of a reference speech. Existing VisualTTS models typically adopt lightweight architectures and design specialized modules to achieve the above goals respectively, yet the speech quality is not satisfied due to the model capacity and the limited data in VisualTTS. Recently, speech large language models (SpeechLLM) show the robust ability to generate high-quality speech. But few work has been done to well leverage temporal cues from video input in generating lip-synchronized speech. To generate both high-quality and lip-synchronized speech in VisualTTS tasks, we propose a novel Visual Speech Language Model called VSpeechLM based upon a SpeechLLM. To capture the synchronization relationship between text and video, we propose a text-video aligner. It first learns fine-grained alignment between phonemes and lip movements, and then outputs an expanded phoneme sequence containing lip-synchronization cues. Next, our proposed SpeechLLM based decoders take the expanded phoneme sequence as input and learns to generate lip-synchronized speech. Extensive experiments demonstrate that our VSpeechLM significantly outperforms previous VisualTTS methods in terms of overall quality, speaker similarity, and synchronization metrics.


\end{abstract}

%% file: 01-intro.tex
\section{Introduction}

\begin{figure}
    \centering
    \includegraphics[width=0.9\linewidth]{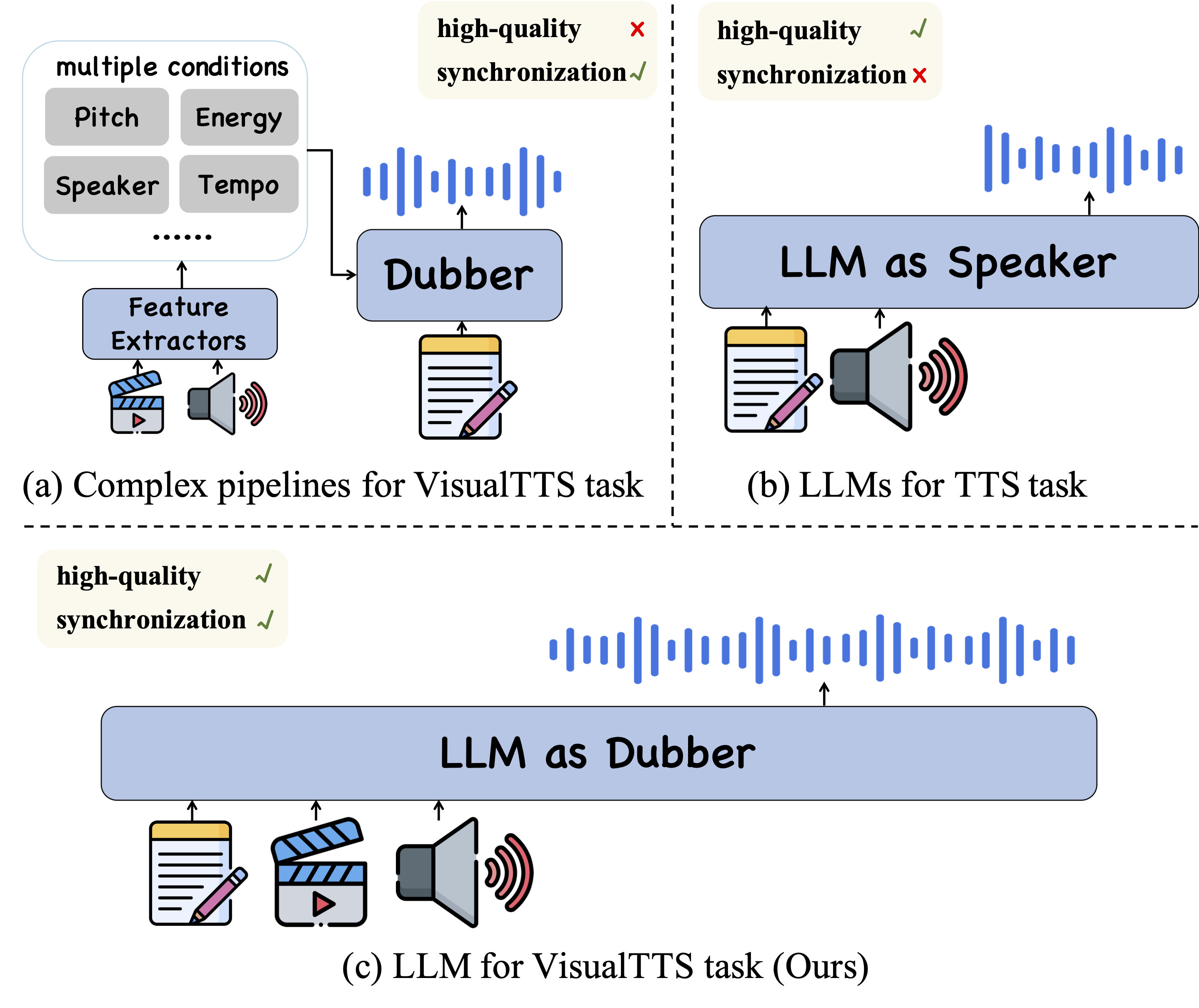}
    \caption{Comparison of our proposed approch with related works. (a) shows the previous VisualTTS method; (b) shows the conventional SpeechLLMs for the TTS task; (c) shows the proposed VSpeechLM designed for the VisualTTS task.}
    \vspace{-2.2 em}
    \label{fig:teaser}
\end{figure}

The Visual Text-to-Speech (VisualTTS) task aims to dub a video by synthesizing speech that aligns with input text, mimics the timbre of a reference voice and synchronizes with the lip movements of a speaker in the video~\cite{tacotron}. It has broad applications in film dubbing, video post-production, and multilingual translation, but remains highly challenging. It requires the model to generate \textbf{high-quality and natural} speech that is easy to understand and highly similar to the reference timbre, and simultaneously to achieve precise \textbf{lip synchronization}, placing higher demands on the model.

Most prior VisualTTS works~\cite{Neural_Dubber, v2c, hpmdubbing, speaker2dub, styledubber, mcdubber, context_icassp,vssflow,t2va,lova} use lightweight TTS backbones~\cite{fastspeech2,matchatts}. These methods (Figure~\ref{fig:teaser}(a)) often use external modules to extract speech-related features, such as speaker embeddings, and design auxiliary modules for speaker enhancement and prosody modeling.  However, these methods struggle to scale due to limited training data and model capacity, facing challenges in generating high-quality speech with the video condition. 

Recently, Speech Large Language Models (SpeechLLMs)~\cite{audiolm, valle,moshi,uniaudio,VoxtLM,SPIRITLM} have made great progress in the TTS task. By discretizing speech into tokens and formulating speech synthesis as a language modeling task, SpeechLLMs effectively leverage large-scale diverse datasets, making them robust to noisy training data and capable of generating high-quality speech.  However, they lack the ability to leverage lip information from the visual modality, making fine-grained synchronization difficult (Figure~\ref{fig:teaser}(b)). For example, DubWise~\cite{dubwise} attempts to incorporate video features into SpeechLLM via cross-attention, but only control global duration without fine temporal alignment. Effectively incorporating temporal cues from the visual modality into SpeechLLMs remains a key challenge.

In this work, we propose VSpeechLM, a novel Visual Speech Language Model that improves both speech quality and lip synchronization in VisualTTS (Figure~\ref{fig:teaser}(c)). We adopt a SpeechLLM as the decoder to ensure high-quality speech generation. To integrate visual cues, we introduce a text-video aligner, which first learns fine-grained temporal alignment between phonemes and lip movements, then expands the phoneme sequence based on the alignment to express the synchronization between text and video. Then the expanded phoneme sequence, combined with reference speech as a prompt, guides the SpeechLLM based decoders to generate speech that is both natural-sounding and lip-synchronized. We evaluate our VSpeechLM on two widely used VisualTTS datasets in terms of overall speech quality, speaker similarity, intelligibility, and duration consistency. Experimental results demonstrate that our approach consistently outperforms previous baselines, validating the effectiveness of our text-video aligner and SpeechLLM integration. Here are our three main contributions:
\begin{itemize}[leftmargin=*]
    \item We propose a VisualTTS framework called VSpeechLM that extends the visual modeling capability of SpeechLLM, capable of generating high-quality and lip-synchronized speech conditioned on input text, video, and reference speech.
    
    \item We propose a text-video aligner that effectively integrate fine-grained visual temporal cues into the SpeechLLM based decoders, helping it generate lip-synchronized speech.
    
    \item Extensive experiments demonstrate that our approach achieves state-of-the-art performance on two widely used VisualTTS datasets across multiple metrics on speech quality, speaker similarity, intelligibility, and synchronization.
   
\end{itemize}

%% file: 02-related.tex
\section{Related Work}

\begin{figure*}[htbp]
    \centering
    \includegraphics[width=1.0\linewidth]{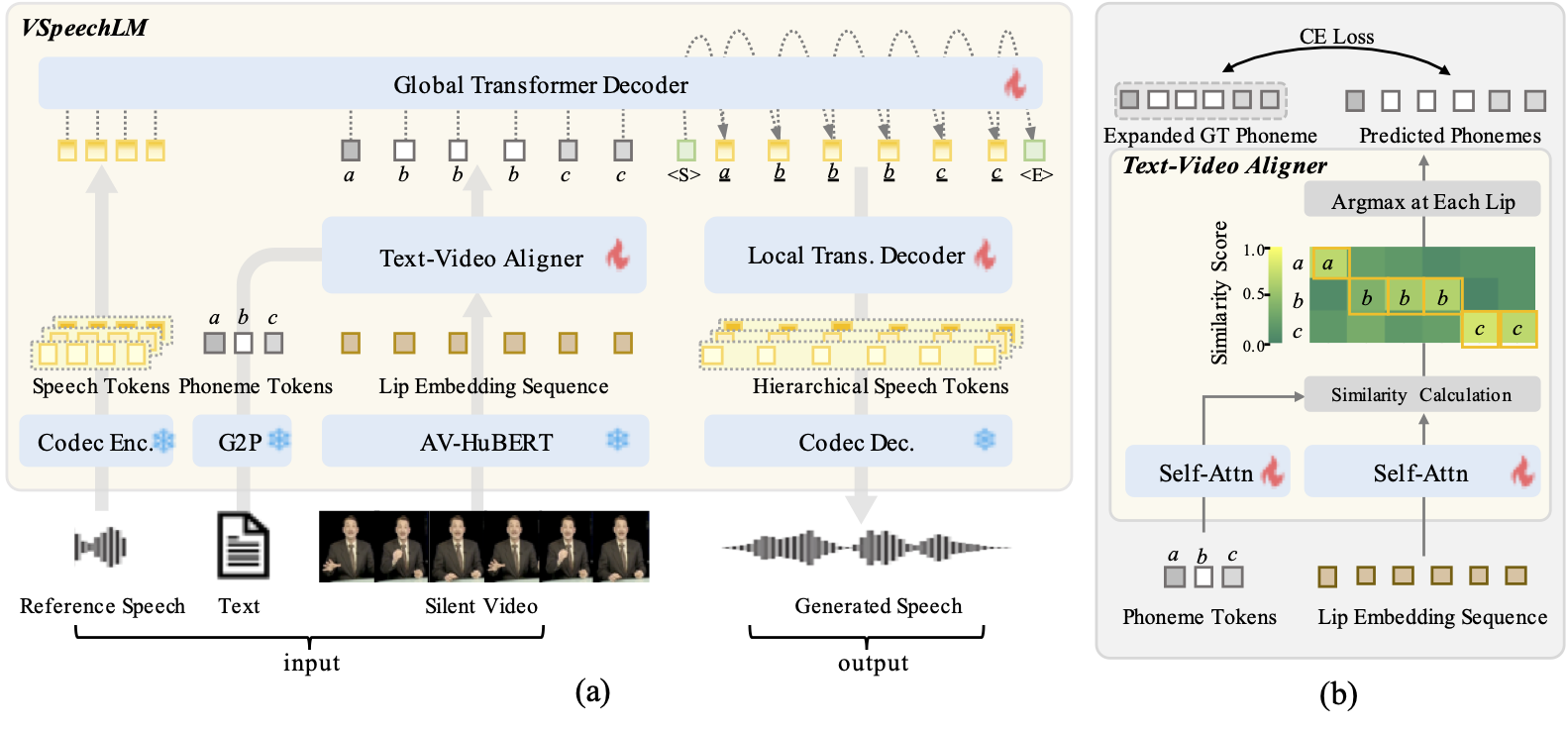}
    \caption{The architecture of our proposed VSpeechLM. (a) shows the overall architecture of VSpeechLM. It is composed of three kinds of components: Feature Extractors that turn the input modalities into representations; a text-video aligner that generates an expanded phoneme sequence by capturing the synchronization relationship between phonemes and lip movements; and a SpeechLLM Decoder that learns to generate lip-synchronized speech. (b) shows the detailed architecture of our proposed text-video aligner.}
    \label{fig:model(1)}
\end{figure*}

\subsection{Visual Text-to-Speech Generation}

Previous VisualTTS models~\cite{hpmdubbing, dsu, speaker2dub, styledubber,emodubber} are primarily based on lightweight TTS backbones such as FastSpeech2~\cite{fastspeech2} or flow-matching TTS models~\cite{matchatts}. These models often rely on speaker embeddings~\cite{hpmdubbing} or dedicated speaker modules~\cite{speaker2dub, styledubber, emodubber}, and require auxiliary features like pitch and energy, which increase architectural complexity. However, their performance is still limited by model capacity and data scale. In this paper, we propose a VisualTTS framework built on a SpeechLLM that inherently produces high-quality speech without the need for complex feature engineering. To achieve fine-grained temporal control, we introduce a text-video aligner that expands the phoneme sequence based on the learned audio-visual alignment, guiding the SpeechLLM to generate lip-synchronized speech.
\vspace{-0.3cm}
\subsection{Speech Large Language Models}
Recently SpeechLLMs show powerful capabilities in speech generation~\cite{audiolm,valle}, formulating it as a language modeling task over discretized speech tokens~\cite{soundstream,encodec,espnet-codec,hubert}. By using phonemes and reference tokens as prompts, these models can be trained on large-scale datasets to generate diverse and natural speech.
Many recent SpeechLLMs~\cite{valle,uniaudio} adopt multi-layer speech codecs~\cite{soundstream,espnet-codec} to represent speech at different granularities, improving quality but increasing sequence length and computation. Many approaches adopt hierarchical modeling to solve this problem: VALL-E~\cite{valle} first generates key tokens from the first quantizer using an autoregressive model and then uses a non-autoregressive model to generate tokens from other layer. UniAudio~\cite{uniaudio} and MoShi~\cite{moshi} use global-local Transformers to reduce cost by decoupling inter- and intra-frame modeling. However, most SpeechLLMs are limited to audio-text inputs and cannot leverage visual information, making them unsuitable for VisualTTS. To address this, we propose a new model called VSpeechLM, which extends the SpeechLLM to incorporate video modality. A text-video aligner provides an expanded phoneme sequence enriched with lip-synchronization cues, guiding the model to generate speech that is both high-quality and precisely synchronized with lip movement.

%% file: 03-method.tex
\section{Our Approach}

\subsection{Overiew}
To generate both high-quality and synchronized speech given a video, a text and a reference voice, we propose VSpeechLM, a Visual Speech Language Model that extends SpeechLLM with temporal alignment capabilities. As shown in Figure~\ref{fig:model(1)}(a), VSpeechLM consists of three components: Modality Representation modules, a text-video aligner, and a SpeechLLM-based Decoder. The \textbf{Modality Representation modules} extract phonemes from text, lip embeddings from video, and discretize reference speech into speech tokens. The \textbf{text-video aligner} models temporal alignment by computing a similarity matrix between phonemes and lip embeddings, expanding the phoneme sequence to match video length, thereby incorporating lip-synchronization cues. The \textbf{SpeechLLM-based Decoder} employs global and local Transformers: the global Transformer generates temporal context vectors aligned with lip frames, while the local Transformer decomposes these vectors into hierarchical speech tokens for decoding into speech. Details on each module and training are provided in the following subsections.

\subsection{Modality Representation}\label{sec:feature} \label{sec:codec}
We detail our representation of the three input modalities. 
\begin{itemize}[leftmargin=*]
    \item \textbf{Text modality}: We use the grapheme-to-phoneme (G2P) tool\footnote{https://github.com/Kyubyong/g2p} to extract the discrete phoneme sequence $\mathbf{P}$ of length $\boldmath{T_p}$.
    \item \textbf{Video modality}: Temporal cues come from lip movements. We extract lip embeddings $\mathbf{L} \in \mathbb{R}^{T_v \times dim}$ using the AV-HuBERT~\cite{avhubert} model, where $T_v$ is the number of input video frames sampled at 25fps, and $dim =1024$ is the embedding dimension.
    \item \textbf{Speech modality}: Since VisualTTS is treated as a language modeling task, we tokenize speech using the high-fidelity discrete speech tokenizer Encodec~\cite{encodec}, to convert speech into hierarchical discrete token sequences. A speech segment $s$ is encoded into a sequence of hierarchical discrete tokens:
    \begin{align}
        C & = [C_1,C_2, ..., C_{T_s}] =  \text{Encoder}_{\text{encodec}}(s) \in \mathbb{N}^{T_s \times N_q}, \\
        C_t & = [C_t^1, C_t^2, ..., C_t^{N_q}]^{\top} \in \mathbb{N}^{Nq \times 1},
    \end{align}
    where $T_s$ is the time steps of the speech sequence, $N_q = 8$ is the number of codebooks in the EnCodec model. $C_t \in \mathbb{N}^{N_q \times 1}$ is the $t$-th step of the speech sequence, and $C_t^i$ is the index selected from the $i$-th codebook at step $t$. The sample rate is 16k Hz. Both reference and target speech are discretized as speech tokens. The reference speech serves as part of the prompt input to VSpeechLM's decoder, providing speaker identity information. The target speech is the model's target output for training.
\end{itemize}

Given the fixed temporal correspondence between video and speech, their sequence lengths can be aligned using repetition-based interpolation or by adjusting the sampling rates. For notational simplicity, we assume that video and speech share the same sequence length (i.e. $T_v = T_s$) in this paper.

\subsection{Text-Video Aligner} \label{sec:aligner}
We propose a text-video aligner that expands a raw phoneme sequence to match the video's lip movements along the time axis, producing an expanded phoneme sequence with duration cues. As shown in Figure~\ref{fig:model(1)}(b), given the phoneme sequence $\mathbf{P}$ and the lip embeddings $\mathbf{L}$, the  aligner computes a \textbf{\textit{similarity matrix}} and assigns a phoneme to each video frame, generating an expanded sequence aligned with lip motion.

\subsubsection{\textbf{Phoneme-Lip Similarity Matrix.}} 
To align the phoneme sequence $\mathbf{P}$ (with length $T_p$) and lip embeddings $\mathbf{L}$ (with length $T_v$), we use two encoders to project them into a shared representation space. The phoneme encoder embeds $\mathbf{P}$ into continuous features $\mathbf{P_{enc}} \in \mathbb{R}^{T_p \times dim}$, and the lip encoder transforms $\mathbf{L}$ into $\mathbf{L_{enc}} \in \mathbb{R}^{T_v \times dim}$, where $dim$ is the hidden dimension of the feature vectors. Both encoders adopt self-attention mechanism, enabling the model to capture complex dependencies by attending to interactions between all positions in the input sequence.

Then we can compute the similarity matrix $\mathbf{A}$ between the two sequences, which serves as the alignment result:
\begin{equation}
    \mathbf{A} = \text{Softmax}(\frac{\mathbf{P_{enc}} \mathbf{L_{enc}}^\top}{\sqrt{dim}}) \in \mathbb{R}^{T_p \times T_v},
\end{equation}
where $T_v$ and $T_p$ denote the length of $\mathbf{L_{enc}}$ and $\mathbf{P_{enc}}$ respectively, and $dim$ denotes the hidden dimension. The element $a_{i,j}$ in $\mathbf{A}$ indicates the similarity between the $i$-th phoneme and the $j$-th lip frame.

\subsubsection{\textbf{Expanding Phoneme according to Phoneme-Lip Similarity.}} \label{sec:expand}
Similarity matrix $\mathbf{A}$ provides temporal alignment scores between phonemes and lip movements. Based on it, we can identify the max similarity score of each column (lip column). Through this, each lip movement is associated with a phoneme, then forms the final \textbf{expanded phoneme sequence}:
\begin{align}
    & \mathbf{P_{exp}} = \text{Repeat}(\mathbf{P}[\text{argmax}(\mathbf{A}, dim=1)]), \mathbf{P_{exp}} \in \mathbb{R}^{T_v}
\end{align}
where $\mathbf{P}$ is the input phoneme sequence, $\mathbf{A}$ is the similarity matrix. The expanded sequence $\mathbf{P_{\text{exp}}}$ is synchronized with lip motion and provides fine-grained duration cues for downstream generation.

\subsection{SpeechLLM based Decoder}\label{sec:decoder}
VSpeechLM's Decoder adopts a multiscale Transformer structure, consisting of a global Transformer and a local Transformer to model speech at different granularities. Unlike prior SpeechLLMs~\cite{uniaudio,moshi}, we feed the global Transformer the expanded phoneme sequence $\mathbf{P_{\text{exp}}}$, enabling it to leverage duration cues and generate a temporal context vector for each step. The local Transformer then decomposes these context vectors into hierarchical speech tokens.

\begin{table*}[htbp]
\scalebox{0.8}{
\begin{tabular}{ccccccccccc}
\toprule
\multicolumn{1}{c|}{} & \multicolumn{5}{c|}{Results on Chem Dataset} & \multicolumn{5}{c}{Results on GRID Dataset} \\ 
\midrule
\multicolumn{1}{l|}{} & WER $\downarrow$ & Spk Sim $\uparrow$ & UTMOS $\uparrow$ & MCD-DTW $\downarrow$ & \multicolumn{1}{c|}{MCD-DTW-SL $\downarrow$} & WER $\downarrow$ & Spk Sim $\uparrow$ & UTMOS $\uparrow$ & MCD-DTW $\downarrow$ & MCD-DTW-SL $\downarrow$ \\ 
\midrule

\multicolumn{1}{l|}{GT} & 3.87 & 100 & 4.19 & 0.00 & \multicolumn{1}{c|}{0.00} & 22.5 & 100 & 4.04 & 0.00 & 0.00 \\
\multicolumn{1}{l|}{GT (decoder)} & 3.89 & 99.8 & 3.77 & 2.60 & \multicolumn{1}{c|}{2.61} & 22.6 & 93.3 & 3.98 & 2.62 & 2.66  \\ 
\midrule
\multicolumn{11}{l}{\textbf{\textit{Baselines}}} \\ 
\midrule
\multicolumn{1}{l|}{DSU (InterSpeech2023)~\cite{dsu}} & 38.6 & 72.9 & 3.33 & 6.69 & \multicolumn{1}{c|}{6.72} & 39.9 & 8.60 & 3.55 & 10.6 & 10.6 \\ 

\multicolumn{1}{l|}{HPMDubbing (CVPR2023)~\cite{hpmdubbing}} & 29.8 & 44.6 & 3.11 & 6.91 & \multicolumn{1}{c|}{8.56} & 44.2 & 31.3 & 2.11 & 6.79 & 7.09 \\

\multicolumn{1}{l|}{StyleDubber (ACL2024)~\cite{styledubber}} & 14.2 & 73.7 & 3.14 & \underline{6.01} & \multicolumn{1}{c|}{\underline{6.36}} & \underline{19.6} & \underline{67.0} & 3.73 & 6.33 & 6.42 \\

\multicolumn{1}{l|}{EmoDubber (CVPR2024)~\cite{emodubber}} & \underline{12.8} & \underline{78.1} & \textbf{3.87} & 6.51 & \multicolumn{1}{c|}{6.51} & 19.8 & 50.5 & \textbf{3.98} & \underline{3.92} & \underline{3.92} \\

\midrule
\multicolumn{11}{l}{\textbf{\textit{Ours}}} \\ 
\midrule
\multicolumn{1}{l|}{VSpeechLM} & \textbf{12.5} & \textbf{78.9} & \underline{3.73} & \textbf{5.16} & \multicolumn{1}{c|}{\textbf{5.28}} & \textbf{18.3} & \textbf{69.7} & \underline{3.94} & \textbf{3.91} & \textbf{3.90} \\ 
\bottomrule
\end{tabular}}
\caption{Comparison of our method with baselines on the datasets Chem and GRID. For WER, MCD-DTW and MCD-DTW-SL, we follow the same computing methods as the baselines~\cite{hpmdubbing,styledubber,emodubber} and compare our results with their reported scores in~\cite{emodubber}. For UTMOS and Spk Sim, we use their public checkpoint to generate speech and evaluate. The highest score for each metric is in bold, while the second highest score is in underlined.\protect\footnotemark}
\vspace{-1 em}
\label{tab:main}
\end{table*}

\subsubsection{\textbf{Global Transformer.}}
As described in Section~\ref{sec:feature}, a speech segment is encoded as a sequence of hierarchical discrete tokens $C=\left \{ C_t^i \right \}_{t=1 \sim T_s}^{i=1\sim N_q}$. To align with the speech token structure, the phoneme input is replicated $N_q$ times. At each time step $t$, the global Transformer generates a \textbf{temporal context vector}:
\begin{equation}
    h_{t} = \text{Trans}_{\text{global}}(C_{\textless t-1}).
\end{equation}
We leverage the global temporal modeling capability of the global Transformer, letting it learn from the lip-synchronized $\mathbf{P_{exp}}$. As a result, the temporal context vectors for step $t$ contain speech information that is aligned with the corresponding lip frame.

\subsubsection{\textbf{Local Transformer.}}
Then, the local Transformer decomposes the temporal context vector $h_{t}$ from the global transformer into $N_q$ hierarchical tokens autoregressively:
\begin{equation}
    C_t^i = \text{Trans}_{\text{local}}(h_{t}, C_{t-1}^{\textless i}),
\end{equation}
where $C_t^i$ is the index of the $i$-th codebook, which captures information from a specific aspect of the speech signal.

\subsection{Training and Inference}
\subsubsection{\textbf{Training objective of text-video aligner.}}\label{sec:traintv}
We model the learning process of the text-video aligner as a classification task, where each lip movement corresponds to a ground-truth phoneme in the dataset. Specifically, the model should identify the phoneme ``class'' that corresponds to each lip movement via the similarity matrix. We first extract the real duration of each phoneme using the Montreal Forced Aligner (MFA) tool~\cite{MFA}. Repetition of each phoneme according to its duration yields the ground truth expanded phoneme sequence $\mathbf{G}$ (length $T_v$). Consequently, we adopt the cross-entropy loss as the training objective:
\begin{equation}
    \mathcal{L}_{CE}^{align} = - \sum_{i}^{T_v} \log \mathbf{A}^{\mathbf{G_{i}}}_i
\end{equation}
where $\mathbf{A}$ denotes the predicted similarity matrix. $\mathbf{A}^{\mathbf{G_{i}}}_i$ can be viewed as the probability of ``classifying'' the $i$-th lip frame to phoneme $\mathbf{G_i}$.

\subsubsection{\textbf{Training objective of SpeechLLM based Decoder.}}
As described in Section~\ref{sec:decoder}, in one generating step, VSpeechLM's decoder uses the local Transformer to autoregressively generate speech tokens, conditioned on global context from the global Transformer. The decoder is trained to minimize cross-entropy loss, guided by the target speech sequence:
\begin{equation}
    \mathcal{L}_{CE}^{dec} = - \sum_{t=1}^{T_s} \sum_{i=1}^{N_q} \log P(C_t^i | C_t^{\textless i}, C_{\textless t}).
\end{equation}

\subsubsection{\textbf{Training and Inference of VSpeechLM}}
In the training stage, the text-video aligner and the SpeechLLM based Decoders are trained jointly. The training is guided by minimizing the sum of $\mathcal{L}_{CE}^{align}$ and $\mathcal{L}_{CE}^{dec}$. During the training stage, we use the ground truth expanded phoneme sequence $\mathbf{G}$ (see Section~\ref{sec:traintv}) as the input of the SpeechLLM based decoder. In the inference stage, the predicted expanded phoneme sequence $\mathbf{P_{exp}}$ is used. 

%% file: 04-exp.tex
\section{Experiments}
\footnotetext{Sentences in the GRID dataset are short and often contain liaison (connected speech), and non-standard pronunciation. For example, ``lay'' may be misrecognized as ``they'', and ``place green'' as ``play screen'' by the ASR model, leading to suboptimal WER even for ground-truth audio.}

\subsection{Experimental Setup}\label{sec:setup}
\textbf{Datasets.}
\noindent
We use two datasets. 1) Single-speaker English chemistry lecture dataset: Chem~\cite{chem}. Following~\cite{emodubber,styledubber,hpmdubbing}, we split it into 6,240 training, 200 validation, and 200 test samples. 2) Multi-speaker English dataset GRID~\cite{grid} with 33 speakers and 1,000 clips of each speaker. Following~\cite{styledubber,emodubber}, we select 100 clips per speaker for testing, resulting in 32,670 training and 900 test samples.

\noindent
\textbf{Baselines.}
We compare against four baselines: DSU~\cite{dsu}, HPMDubbing~\cite{hpmdubbing}, StyleDubber~\cite{styledubber}, and EmoDubber~\cite{emodubber}. The first three use FastSpeech2-like decoders, while EmoDubber adopts a flow-matching decoder. Details are in the supplementary.

\noindent
\textbf{Metrics.} We evaluate generated speech using five objective metrics: Word Error Rate (WER) for intelligibility, speaker similarity (Spk Sim) for timbre consistency, UTMOS~\cite{utmos} for overall quality, Mel Cepstral Distortion with Dynamic Time Warping (MCD-DTW)\cite{v2c} for acoustic fidelity, and MCD-DTW weighted by speech length (MCD-DTW-SL)\cite{v2c} to assess duration synchronization. Details are in the supplementary.

\noindent
\textbf{Implementation Details.}
We initialize VSpeechLM decoders from ESPnet~\cite{espnet, espnetlm} pretrained models and jointly train all modules. We use AdamW with an initial learning rate of $1\mathrm{e}{-4}$ and betas of (0.9, 0.95). During inference, we apply top-$k$ sampling ($k=30$). The temperature is set to 1.2 for the expressive Chem dataset and 0.5 for the less variable GRID dataset. More model architecture and training details are provided in the supplementary material.

\subsection{Objective Evaluation} 
We compare the performance of VSpeechLM and the baselines on two widely used datasets: Chem and GRID. We report the results in Table~\ref{tab:main}. The left and right sides of the table are the results of Chem and GRID, respectively. Our VSpeechLM exceeds the baselines in all metrics except for UTMOS in the two datasets. Specifically, in terms of intelligibility, as shown by the decrease in WER, VSpeechLM performs better than the SOTA model EmoDubber~\cite{emodubber}. In speaker similarity (Spk Sim), VSpeechLM outperforms other models despite not using an external encoder for speaker embedding or a dedicated speaker modeling module. This demonstrates the strong in-context learning capability of our model. In terms of synchronization (MCD-DTW-SL), on the Chem and GRID datasets, there are improvements of 17.6\% (vs. EmoDubber) and 60.9\% (vs. StyleDubber) reduction respectively compared to the SOTA results, which indicates that our method can achieve better duration consistency, proving the effectiveness of the temporal controlling. In terms of UTMOS, our approach performs the second best, just behind EmoDubber a bit. But it is close to the codec decoder upper bound (only 0.04)—suggesting codec quality may limit further gains. We plan to explore better codecs or direct mel-based reconstruction in the future.

\subsection{Subjective Evaluation} \label{sec:human}
\begin{figure}
    \centering
    \includegraphics[width=1.0\linewidth]{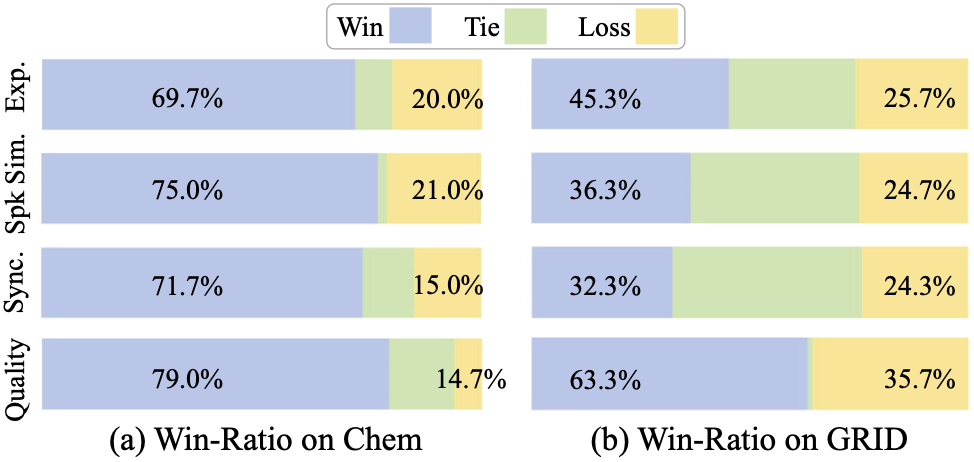}
    \caption{Human Evaluation Results. We compare VSpeechLM with the strongest baselines on Chem and GRID in terms of speech quality (Quality), lip synchronization (Sync), speaker similarity (Spk Sim), and expressiveness (Exp). (a) shows win-tie-loss ratios against EmoDubber on Chem, and (b) against StyleDubber on GRID.}
    \label{fig:human}
\end{figure}

We further conduct subjective evaluation. We select the methods that perform best among baselines on two datasets for comparison. Specifically, we compared our method with EmoDubber on the Chem dataset and with StyleDubber on the GRID dataset. We sampled 50 clips each from the test sets of Chem and GRID. Ten experts are invited to score the models based on four criteria: speech quality (Quality.), lip synchronization (Sync.), speaker similarity (Spk Sim.), and expressiveness (Exp.). The scoring was done by selecting a winner or declaring a tie for each comparison. The results, summarized in Figure~\ref{fig:human}, show that our model outperforms the baselines over four criteria on the two datasets. We find that, when compared to the GRID dataset, there are many ties in Exp. and Sync.. This is because the speech in the GRID dataset was recorded in a studio background, with short sentences and neutral emotions. In this dataset, lip synchronization is relatively easier to capture, making it hard for humans to tell the difference. 

\subsection{Case Studies}
\begin{figure}
    \centering
    \includegraphics[width=1.0\linewidth]{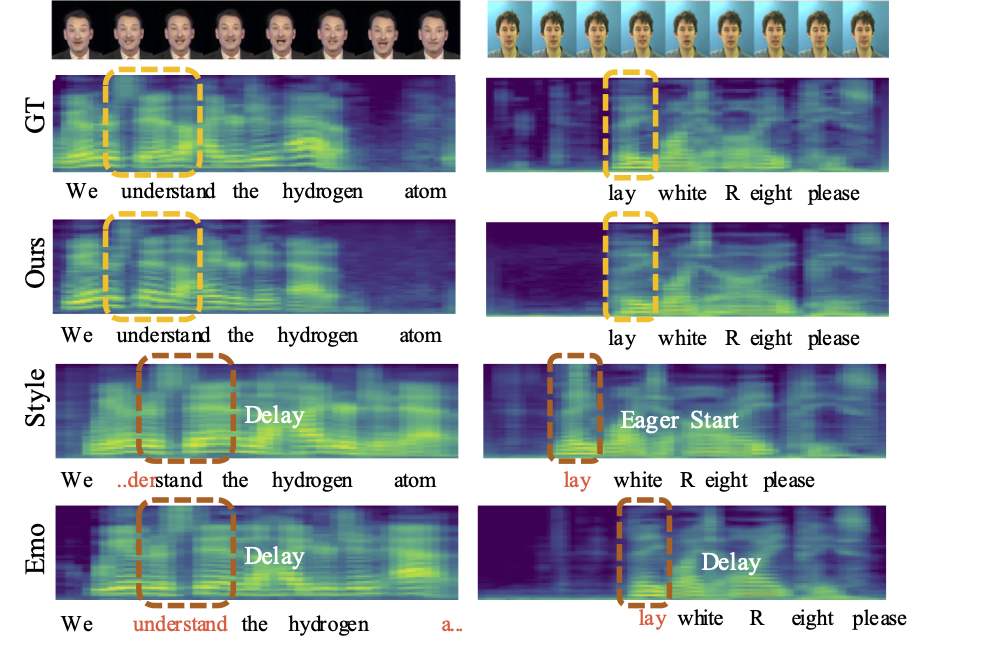} 
    \caption{Case studies on the Chem and GRID test sets. The top two rows show video frames and ground-truth Mel-spectrograms, followed by outputs from VSpeechLM, StyleDubber, and EmoDubber. Mel-spectrograms, with time and frequency represented on the horizontal and vertical axes respectively, allows visual comparison of synchronization and acoustic quality against the ground truth.}
    \label{fig:case}
\end{figure}

Figure~\ref{fig:case} presents two representative cases from both the Chem and GRID test set, demonstrating our VSpeechLM achieves superior synchronization while maintaining high acoustic quality. We compare our VSpeechLM with two baselines that perform better in objective evaluation: StyleDubber and EmoDubber. In Case 1, the two baselines exhibit noticeable delays when pronouncing the word ``understand''. In Case 2, StyleDubber exhibits an early onset when pronouncing the word ``lay'', while EmoDubber shows a delayed onset. In contrast, our proposed VSpeechLM consistently generates clear and well-structured Mel-spectrograms, with speech that is better aligned with ground truth. 

\subsection{Ablations and Analyses}
In this section, we present the main ablation studies and analyses of the model's capabilities. Additional results and analyses can be found in the supplementary material.

\subsubsection{\textbf{
The text-video aligner helps SpeechLLMs based decoders to ``see'' better.
}} \label{exp:alignment}

\begin{table}[]
\scalebox{0.77}{
\begin{tabular}{l|cccc}
\toprule
 & WER $\downarrow$ & Spk Sim $\uparrow$ & UTMOS $\uparrow$ & MCD-DTW-SL $\downarrow$ \\ \midrule
(a) text-video aligner & \textbf{12.48} & \textbf{78.87} & 3.73 & \textbf{5.28} \\
(b) no visual & 15.02 & 76.91 & \textbf{3.81} & 6.49  \\
(c) visual prefix  & 15.16 & 74.38 & 3.51 & 6.94  \\ \bottomrule
\end{tabular}}
\caption{Ablating visual integration methods on the Chem dataset. `No visual' denotes a TTS way without visual information. `visual prefix' denotes a VisualTTS way that visual features are input as a prefix prompt.}
\vspace{-1 em}
\label{tab:visual}
\end{table}

To investigate how to effectively extend visual processing capabilities for SpeechLLMs, we conduct a comparative analysis of three approaches: (a) incorporating visual input via our proposed text-video aligner; (b) performing TTS with the backbone SpeechLLM decoders without any visual input; (c) extending the backbone SpeechLLM decoders with a visual prefix prompt for VisualTTS. We evaluate the three methods on the Chem dataset, with results summarized in Table~\ref{tab:visual}. Our proposed text-video aligner achieves the best MCD-DTW-SL score, demonstrating its effectiveness in facilitating lip-synchronized speech generation. In contrast, method (c) performs the worst in terms of synchronization, indicating that simply concatenating the video as a prompt has a negative effect. This is perhaps because the decoder was not pretrained to process visual inputs, and the limited training data available for VisualTTS is insufficient for the model to adapt to the new visual modality, leading to confusion. In comparison, our approach preserves the decoder’s original input space, making it easier for the model to learn this synchronization pattern effectively.

\subsubsection{\textbf{Do we really need pretraining and training?}} As described in Section~\ref{sec:setup}, the VSpeechLM decoder is initialized with parameters from a pretrained SpeechLM prior to training. To assess the necessity of each training stage, we compare the following three settings: (a) Ours: Our full pipeline, where VSpeechLM decoder is both pretrained and adapted through VisualTTS training; (b) w/o pretraining: VisualTTS training without decoder pretraining; (c) w/o adapting: the pretrained decoder is kept frozen when adapted through VisualTTS training. The results are shown in the Table~\ref{tab:train}. Omitting either decoder pretraining or VisualTTS adaptation leads to a significant performance drop. We attribute this to the pretrained decoder providing a foundational ability to convert text into speech units, while VisualTTS training allows the model to adapt to the extended phoneme sequence. These findings validate the necessity of both components in our proposed pipeline.
\begin{table}[]
\scalebox{0.77}{
\begin{tabular}{l|cccc}
\toprule
 & WER $\downarrow$ & Spk Sim $\uparrow$ & UTMOS $\uparrow$ & MCD-DTW-SL $\downarrow$ \\ \midrule
(a) Ours & \textbf{12.5} & \textbf{78.9} & \textbf{3.87} & \textbf{5.28} \\
(b) w/o pretraining & 151 & 50.6 & 1.62 & 13.99  \\
(c) w/o adapting & 829 & 2.86 & 1.49 & 80.92  \\ 
\bottomrule
\end{tabular}}
\caption{Comparison Under Different Training Configurations. `w/o pretraining' indicates that the decoder is randomly initialized and trained only with VisualTTS. `w/o adapting' indicates that the decoder is pretrained but kept frozen, with only the text-video aligner adapted in the VisualTTS stage.}
\vspace{-1 em}
\label{tab:train}
\end{table}

\subsubsection{\textbf{Can VSpeechLM be generalized to unseen data?}} Due to the domain limitations of the Chem and GRID datasets, we conducted zero-shot experiments on 200 randomly sampled clips from the LRS2~\cite{lrs2} dataset to assess the model's generalization ability on complex and diverse data. The LRS2 dataset contains spoken sentences from BBC television, featuring background noise and non-frontal face poses. The results are summarized in Table~\ref{tab:lrs2}. Our model achieves the best performance on LRS2, demonstrating robustness under challenging real-world conditions.

\begin{table}[]
\scalebox{0.77}{
\begin{tabular}{l|cccc}
\toprule
 & WER $\downarrow$ & Spk Sim $\uparrow$ & UTMOS $\uparrow$ & MCD-DTW-SL $\downarrow$ \\ \midrule
GT & 8.52 & 1.00 & 3.14 & 0.00 \\
HPMDubbing & 198 & 33.2 & 1.28 & 8.43  \\
StyleDubber & 91.4 & 50.3 & 1.87 & 14.0  \\ 
EmoDubber & 110 & 10.3 & 1.70 & 8.68  \\ 
VSpeechLM & \textbf{16.1} & \textbf{53.9} & \textbf{2.98} & \textbf{7.59}  \\ 
\bottomrule
\end{tabular}}
\caption{Zero-shot evaluation results of VSpeechLM on 200 randomly sampled clips from the LRS2 dataset. LRS2 contains noisy audio and non-frontal face poses, presenting challenging real-world conditions.}
\vspace{-1 em}
\label{tab:lrs2}
\end{table}

%% file: 05-conclusion.tex
\section{Conclusion and Future Work}

In this work, we propose a novel Visual Speech Language Model (VSpeechLM) based on a SpeechLLM for generating high-quality and lip-synchronized speech. To effectively integrate the temporal cues provided by the visual modality into the conventional SpeechLLM, we design a text-video aligner. It provides an expanded phoneme sequence containing lip-synchronization cues. The SpeechLLM based decoders take this expanded sequence as a prompt and learns to generate lip-synchronized speech. Experimental results demonstrate that our proposed VSpeechLM achieves strong performance across multiple evaluation metrics. 
In future work, we aim to advance the capabilities of VisualTTS generation by addressing more intricate and challenging scenarios, such as generating coherent speech when facing partially invisible lip motion or low-quality video inputs.